\documentclass[doublecol]{epl2} 
\title{Robust unidirectional transport in a one-dimensional metacrystal with long-range hopping}
\shorttitle{Robust unidirectional transport in ... } %Insert here a short version of the title if it exceeds 70 characters

\author{S. Longhi \inst{1,2}}
\shortauthor{S. Longhi}

\institute{                    
  \inst{1}  Dipartimento di Fisica, Politecnico di Milano, Piazza L. da Vinci 32, I-20133 Milano, Italy\\
  \inst{2}  Istituto di Fotonica e Nanotecnlogie del Consiglio Nazionale delle Ricerche, sezione di Milano, Piazza L. da Vinci 32, I-20133 Milano, Italy}

\abstract{ In two- and three-dimensional structures,  topologically-protected chiral edge modes offer a powerful mean to realize robust light transport. However, little attention has been paid so far to robust one-way transport in one-dimensional systems. Here it is shown that unidirectional transport, which is immune to disorder and backscattering, can occur in certain one-dimensional metacrystals with long-range hopping without resorting to topological protection. Such metacrystals are described by an effective Hermitian Hamiltonian with broken time reversal symmetry, and transport does not require adiabatic (Thouless) pumping. A simple implementation in optics of such one-dimensional metacrystals, based on transverse light dynamics in a self-imaging optical cavity with phase gratings, is suggested.}
\begin{document}

\maketitle

\section{Introduction}
Topological photonic structures, a new class of optical systems inspired by quantum
Hall effect and topological insulators, have attracted a huge attention in recent years owing to their rather unique property of
permitting robust transport via topologically-protected chiral edge modes \cite{r1}. Such two-dimensional (2D) or three-dimensional (3D) optical structures are 
usually realized by breaking time-reversal symmetry, e.g. using magneto-optic media \cite{r2,r3,r4,r5,r6,r7}, or by the introduction of synthetic gauge fields \cite{r8,r9,r10,r11,r12,r13,r14,r15,r16,r17}. Other examples of chiral edge transport in 2D or 3D optical media include photonic Floquet topological insulators in helical waveguide lattices \cite{r18,r18bis}, gyroid photonic crystals \cite{r19},  bianisotropic metamaterials \cite{r20,r21}, chiral hyperbolic metamaterials \cite{r22}, and optomechanical  lattices \cite{r23}. In one-dimensional (1D) systems, the possibility of realizing robust one-way transport has received less attention so far, mainly because topological protection is generally unlikely in 1D. Proposals include 'Thouless pumping' in quasicrystals \cite{r24,r25}, Landau-Zener transport in binary lattices \cite{r26} ,the use of 'synthetic' dimensions in addition to the physical spatial dimension \cite{r27,r28,r28bis}, and non-Hermitian transport \cite{r29}. 
In 1D lattices with short-range hopping, the action of synthetic gauge fields is generally trivial, as any loop encloses zero flux, and thus topological protection can arise from the adiabatic change of some parameter (Thouless pumping) or by adding a 'synthetic' dimension. However, 2D lattices can be mapped into 1D chains with long-range hoppings \cite{r30},  so that loops with nonvanishing magnetic fluxes and quantum Hall physics become possible even in 1D systems without additional synthetic dimensions \cite{r31}. An implementation of 1D lattices with long-range hopping and synthetic gauge fields, based on periodically-driven spin chains with special driving protocols, has been recently suggested in Ref.\cite{r31}. However, its practical realization remains challenging. \par
In they Letter it is shown that in a wide class of 1D metacrystals, described by an effective Hermitian Hamiltonian with long-range hopping and broken time reversal symmetry, one can realize unidirectional and robust transport which is not assisted by topological protection. A simple physical implementation of such metacrystals in optics, based on transverse light dynamics in a self-imaging optical resonator with phase gratings, is suggested.

\section{Robust unidirectional transport in a one-dimensional metacrystal}  

Let us consider the motion of a quantum particle on a 1D lattice, subjected to an external potential $U(x)$ which varies slowly over the lattice period $a$. In our analysis, the potential $U(x)$ accounts for lattice defects or disorder of site energies.  In the single band approximation and after expanding the wave function $\psi(x,t)$ of the particle on the basis of displaced Wannier functions $W(x-na)$, i.e. after setting $\psi(x,t)=\sum_n f(n,t)W(x-na)$, it is well-known that the envelope function $f(n,t)$ is obtained as $f(n,t)=\phi(x=na,t)$, where $\phi(x,t)$ satisfies the Schr\"{o}dinger equation (taking $\hbar=1$)
\begin{equation}
 i \partial_t \phi= \hat{H}_{eff} \phi
 \end{equation}
 with an effective Hamiltonian $\hat{H}_{eff}=\hat{H}_{0}+U(x)$ \cite{r32}, where
\begin{equation}
\hat{H}_0=E(-i \partial_x),
\end{equation}
$E(k)=E(k+ 2 \pi/a)$ is the energy dispersion curve of the lattice band, and $-\pi/a\leq k < \pi/a$ is the Bloch wave number (quasi-momentum).
After introduction of the Fourier coefficients $J_n$ of $E(k)$, $E(k)=\sum_n J_n \exp(inak)$, the Sch\"{o}dinger equation (1) reads explicitly
\begin{equation}
i \frac{\partial \phi (x,t)}{\partial t} =\sum_n \ J_n \phi(x+na,t)+U(x) \phi(x,t).
\end{equation}
Note that $J_n$ corresponds to the hopping amplitude between two sites in the lattice spaced by $n$. For an Hermitian lattice with time-reversal symmetry, the energy $E(k)$ is real and with the even symmetry $E(-k)=E(k)$, which implies $J_n$ real and $J_{-n}=J_n$. For example, in the nearest-neighbor tight-binding approximation (short-range hopping), $E(k) =-J \cos(ka)$, where $J/2=-J_1$ is the hopping amplitude between adjacent lattice sites. The even symmetry of the dispersion curve $E(k)$ is responsible for backscattering of a particle wave packet, propagating along the lattice, in the presence of defects or disorder. In fact, for a forward-propagating wave packet with carrier quasi-momentum $k_0$, moving with a group velocity $v_g=(dE/dk)_{k_0}>0$, the scattering potential can excite the energy-degenerate state with quasi-momentum $-k_0$, corresponding to a backward propagating wave $(dE/dk)_{-k_0}<0$; see Figs.1(a) and (b). By breaking time-reversal symmetry, one can in principle synthesize a lattice band with a dispersion curve $E(k)$ which is an increasing (or decreasing) function of $k$ over the entire Brillouin zone $-\pi/a<k<\pi/a$, with a rapid (abrupt) change at the Brilluoin zone edges $k= \pm \pi/a$; see Fig.1(c). In this way, back reflections are forbidden, since at any quasi-momentum $k_0$ the group velocity has the same sign (apart from the Brillouin zone edges of negligible measure). In such a metacrystal, long-range hopping is necessary, together with a proper engineering of the phases of hopping amplitudes to break time reversal symmetry. For example, in a metacrystal with a sawtooth-shaped dispersion curve of bandwidth $2J$,
\begin{equation}
E(k) =J a k/ \pi, \; \; \; -\pi/a <k < \pi/a,
\end{equation}
the hoping amplitudes $J_n$, as obtained from the Fourier series expansion $E(k)=2J \sum_{n=1}^{\infty} [(-1)^{n+1}/n \pi] \sin (nka)$, are given by
\begin{eqnarray}
J_n= \left\{
\begin{array}{cc}
 0 & n=0 \\
 (-1)^{n+1} J/(\pi i n) & n \neq 0 \\
 %-(-1)^{|n|+1} J / ( 2 \pi i |n|) & n \leq -1.
\end{array}
\right.
\end{eqnarray}
Note that, while $J_{-n} =J_n^*$ (Hermitian lattice), $J_n$ is imaginary, indicating that time reversal symmetry is broken. An interesting property of the sawtooth metacrystal is that, besides of ensuring one-way propagative states, the group velocity $v_g=Ja/ \pi$ is uniform, corresponding to vanishing of group velocity dispersion and distortionless  wave packet propagation. 
To highlight  the robust propagation properties of the sawtooth metacrystal as compared to an ordinary tight-binding crystal with short-range hopping and time reversal symmetry, described by a sinusoidal band $E(k)=- J \cos (ka)$, we numerically-computed the evolution of a Gaussian wave packet in two lattices in the presence of either a potential site defect [Fig.2(a)] and site-energy disorder [Fig.2(b)], assuming the same bandwidth $2J$ and lattice period $a$. The figure clearly shows that, while back reflections and deceleration of motion is observed in the sinusoidal lattice band, wave packet propagation turns out to be robust in the sawtooth metacrystal. 

\begin{figure}
\includegraphics[width=8.4cm]{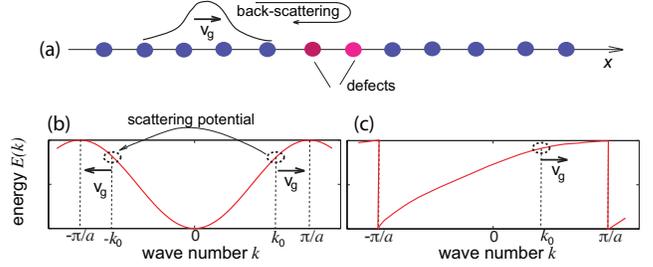}
\caption{ (Color online)  (a) Schematic of back-reflection induced by defects or disorder in a single-band tight-binding lattice. (b) Band dispersion curve $E(k)$ in a crystal with time-reversal symmetry. A forward-propagating wave packet with carrier wave number  $k_0$ can be scattered off by the defects into a backward propagating wave packet with carrier wave number $-k_0$.  (c) Band dispersion curve $E(k)$ in a metacrystal with broken time reversal symmetry and with $dE/dk>0$ over the entire Brillouin zone ($ k \neq \pm \pi/a)$.}
\end{figure}

 \begin{figure}[htb]
\centerline{\includegraphics[width=9.1cm]{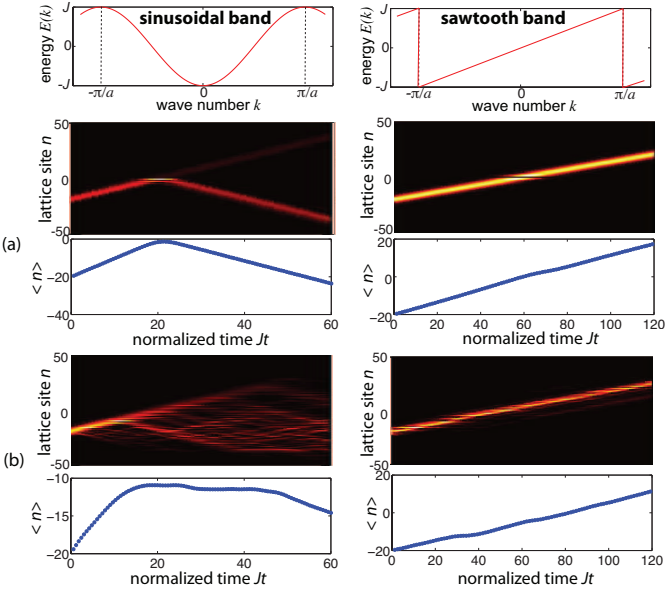}} \caption{ 
(Color online) Propagation of a Gaussian wave packet in a tight-binding crystal with unbroken time reversal symmetry (sinusoidal band, left column)  and with broken time reversal symmetry (sawtooth band, right column). The panels show the numerically-computed temporal evolution of $|f(n,t)|^2$ (on a pseudocolor map) and of the wave packet center of mass $\langle n \rangle$. The initial condition is $f(n,0) \propto \exp[-(n+20)^2/16+i \pi n/2]$, corresponding to a group velocity $v_g=Ja$ in the sinusoidal crystal, and $v_g=Ja/ \pi$ in the sawtooth metacrystal. In (a) a potential defect at site $n=0$ is introduced, namely $U(x=na)=U_0 \delta_{n,0}$ with $U_0=2J$. In (b) on-site potential energy disorder is introduced, corresponding to $U(x=na)$ random variable uniformly distributed in the range $(-J/2,J/2)$.}
\end{figure}

 \begin{figure}[htb]
\centerline{\includegraphics[width=8.4cm]{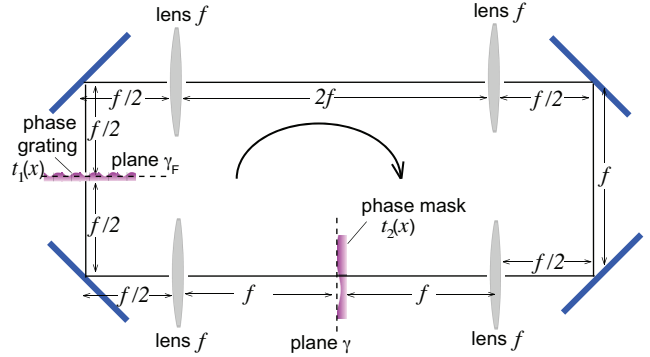}} \caption{ 
(Color online)  Schematic of the optical ring resonator in the self-imaging regime that realizes a metacrystal. The resonator is composed by four lenses (focal length $f$), a phase grating and a phase mask, with transmission $t_1(x)=\exp[-i \varphi_1(x)]$ and $t_2(x)=\exp[-i \varphi_2(x)]$, respectively.}
\end{figure}

\section{Resonator optics realization of a metacrystal} A main challenging is the physical implementation of a 1D metacrystal, which requires long-range hopping and breaking of time reversal symmetry. A possible platform is provided, at least in principle, by spin chains and trapped ions with synthetic gauge fields \cite{r31}. However, the precise tailoring of hopping rates in amplitude and phase remains a rather challenging task. Here we suggest a rather simple optical implementation of a 1D metacrystal, which is based on transverse beam dynamics in a self-imaging optical resonator with a phase grating. In a few recent works, it has been suggested that light waves propagating back and forth in an optical resonator can  emulate synthetic magnetism \cite{r16,r17} and can realize diffraction management \cite{r33,r34}, i.e. the optical analogue of kinetic energy operator management. Here we show that a self-imaging ring resonator with an intracavity phase grating can emulate for light waves the effective Schr\"{o}dinger equation (3) of a metacrystal. A schematic of the passive optical resonator is shown in Fig.3. It consists of a four-lens ring cavity of total length $L=8f$ in the so-called 4-$f$ self-imaging configuration \cite{r34}.  Planes $\gamma$ and $\gamma_F$ shown in Fig.3 are Fourier conjugate planes. A thin phase grating with spatial period $A$ and transmission function $t_{1}(x)=\exp[-i \varphi_1(x)]$, with $\varphi_1(x+A)=\varphi_1(x)$, is placed at the Fourier plane $\gamma_F$, whereas a second phase mask with transmission function $t_{2}(x)=\exp[-i \varphi_2(x)]$ is placed at the plane $\gamma$. Light propagation inside the optical ring can be readily obtained by application of the generalized Huygens-Fresnel integral. Assuming one transverse spatial dimension $x$ and disregarding at this stage of the analysis cavity losses and external beam injection, the evolution of the intracavity field envelope  $\psi_m(x)$ at plane $\gamma$ in the cavity and at the $m-th$ round trip is governed by the following map 
 \begin{equation}
 \psi_{m+1}(x)= t_2(x) t_1\left(  \frac{i \lambda f}{2 \pi} \frac{\partial}{\partial x} \right) \psi_m(x)
 \end{equation} 
 where $\lambda$ is the wavelength of the circulating optical field. The recurrence relation (6) can be transformed into a Schr\"{o}dinger-like wave equation using a rather standard method \cite{r34}. In the limit $|\varphi_{1,2}(x) | \ll 1$, after first-order expansion $t_{1,2}(x) \simeq 1-i \varphi_{1,2}(x)$ and continuation of the round trip number ($m \rightarrow t$), from Eq.(6) one can derive the following evolution equation for the intracavity field $\psi(x,t)$ at plane $\gamma$
 \begin{equation}
 i \frac{\partial \psi}{ \partial t}= E(-i \partial_x) \psi+ U(x) \psi
  \end{equation}
 where $t$ is the temporal variable in units of the cavity round trip time, and where we has set
 \begin{equation}
 U(x)=\varphi_2(x) \; , \;\;\; E(k)= \varphi_1 \left( -\frac{\lambda f k}{2 \pi} \right).
 \end{equation}
 
  \begin{figure}[htb]
\centerline{\includegraphics[width=9cm]{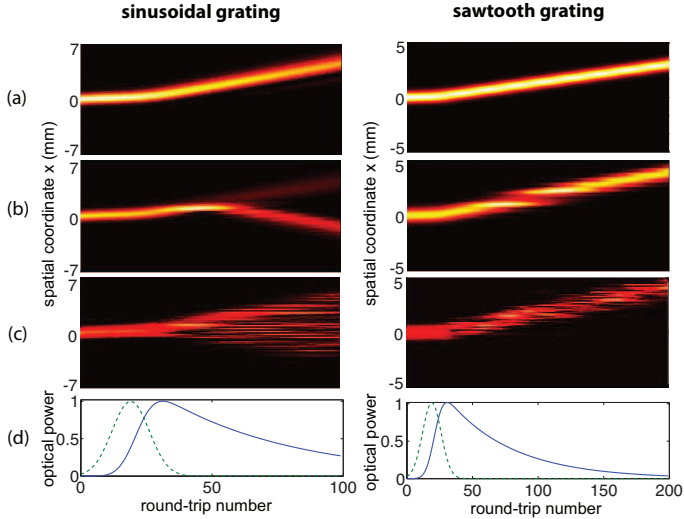}} \caption{ \small
(Color online)  Beam evolution at successive round trips (color maps of normalized intensity
distribution at plane $\gamma$) in the ring resonator with an injected Gaussian pulsed beam for 
a sinusoidal phase grating (left panels) and a for a sawtooth phase grating (right panels) at the Fourier plane $\gamma_F$.
In (a) there is not any phase mask in $\gamma$ (homogeneous metacrystal), whereas on (b) and (c) a phase mask is introduced to emulate 
a localized defect [in (b)] and site energy disorder [in (c)].
Parameter values are given in the text. Panel (d) shows
the behavior of the normalized optical power of the intracavity field (solid line) and the
temporal amplitude $F(t)$ of the Gaussian excitation beam (dashed line).}
\end{figure}

 Note that Eq.(7) is precisely the Schr\"{o}dinger equation (1) of a metacrystal with band dispersion curve $E(k)$ and external potential $U(x)$, defined by Eq.(8). Hence, the transverse beam motion along the spatial coordinate $x$ at the resonator plane $\gamma$ emulates the motion of a quantum particle in an arbitrary 1D crystal under an external potential. The profile of the phase grating in the Fourier plane $\gamma_F$ defines the dispersion curve $E(k)$ of the lattice band, and can be thus tailored to realize a metacrystal, i.e. a crystal with long-range hopping and broken time reversal symmetry. Remarkably, long-range hopping and breaking of time reversal symmetry does not require here to introduce synthetic gauge fields nor special modulation of parameters \cite{r31}, making the method rather simple. The profile of the phase mask at the plane $\gamma$ determines the external potential $U(x)$, and can be designed to emulate lattice defects or disorder. Note that the equivalent spatial period $a$ of the metacrystal in real space $x$ is given by $a= \lambda f /A$. Since the cavity operates in a self-imaging condition and the phase mask and grating act on the $x$ spatial coordinate solely, in the orthogonal $y$ transverse coordinate the beam profile is not affected by propagation inside the resonator.\\ 
 In an optical experiment, wave packet dynamics and robustness against back reflections can be observed by considering the freely-decaying beam dynamics in the passive resonator  initially loaded with a pulsed Gaussian beam $E(x,t)=F(t)G(x)$, which is injected through one of the four cavity mirrors. Assuming that the carrier frequency of the injected beam is in resonance with one of the cavity axial modes, the map (6) is replaced by the following one
 \begin{eqnarray}
  \psi_{m+1}(x) & = & t_2(x) t_1\left(  \frac{i \lambda f}{2 \pi} \frac{\partial}{\partial x} \right)  \psi_m(x)+\sqrt{T} E_m(x)  \nonumber \\
  & - & \frac{T}{2} \psi_m(x),
 \end{eqnarray}
 where $T \ll1$ is the transmittance of the coupling mirror, and $E_m(x)=F(m)G(x)$ is the spatial profile of the injected beam at plane $\gamma$ and at the $m$-th round trip. The free-decay of light in the cavity, following the pulse excitation with the external beam, basically emulates wave packet evolution in a metacrystal. In an experiment, transverse light evolution at successive transits in the cavity can be detected by time-resolved beam profile measurements using a gated camera, as demonstrated e.g. in Refs.\cite{r35,r36}. As an example, Fig.4 shows the  beam evolution of the intracavity field at plane $\gamma$, as obtained by numerical integration of the map (9), assuming either a sinusoidal grating profile $\varphi_1(x)=-J \cos(2 \pi x/A)$, or a sawtooth  grating profile with the same period $A=30 \; \mu$m and amplitude $J=0.5$. Parameter values used in the simulations are $\lambda=633$ nm, $f=2$ cm, and $T=2 \%$, corresponding to a spatial period $a= \lambda f /A = 422 \; \mu$m of the metacrystal. The injected field is a pulsed and tilted Gaussian beam with transverse profile (at plane $\gamma$) $G(x)=\exp(-x^2/w^2+0.5 i \pi x/a)$ ($w=800 \; \mu$m) and pulse envelope $F(t)=\exp[-(t-t_0)^2/\tau^2]$ ($t_0=20$, $\tau=10$ in units of the round trip time $T_R=8f/c \simeq 0.53 $ ns). The external potential $U(x)$ is assumed to be either a localized defect of the lattice ($U(x)=-U_0 \exp[-(x-d)^2/s^2]$, $U_0=0.2$, $d=1600 \; \mu$m, $s=600 \; \mu$m), Fig.4(b); or a random potential ($U(x)$ random variable with uniform distribution in the range $ (-0.5,0.5)$), Fig.4(c). 
 The freely-evolving optical beam, in the absence of lattice defects and disorder, is shown for comparison in Fig.4(a). The behavior of the intracavity power before and after injection with the external beam is also shown in Fig.4(d). The decay of the optical power in the cavity after initial pulse excitation is due to cavity losses at the output coupler. The numerical results clearly indicate that, after excitation of the passive cavity with the tilted external pulsed beam,  transverse beam propagation is robust in the case of the sawtooth phase grating, while back reflections are well visible in the case of the sinusoidal phase grating according to the scenario of Fig.2. 

\section {Conclusions} Robust unidirectional transport can occur in a wide class of 1D Hermitian metacrystals with engineered lattice band. However,  long-range hopping and broken time reversal symmetry are needed to implement such metacrystals. While 1D matter-wave systems, such as spin chains or trapped ions, could be a potential platform to  implement long-range hopping and synthetic gauge fields \cite{r31}, their experimental realization remains challenging. Here we have shown that transverse beam dynamics in a self-imaging optical resonator with a phase grating provides a rather simple and experimentally-accessible system in optics to implement a metacrystal, in which time reversal symmetry breaking and long-range hopping are readily realized without the need for synthetic gauge fields nor special modulation of parameters. The present results disclose an important strategy to realize robust transport in 1D lattices, without resorting to adiabatic (Thouless) pumping \cite{r23,r24} or  non-Hermitian transport \cite{r29}, and suggest resonator optics as a suitable platform to implement a metacrystal.

%\begin{figure}
%\onefigure[width=8.6cm]{Fig1.eps}
%\caption{(Color online) (a) Schematic of a generic Fabry-Perot optical resonator with off-axis longitudinal pumping that realizes the $\mathcal{PT}$-symmetric quantum harmonic oscillator. $ABCD$ is the resonator round-trip matrix with respect to the plane $\gamma$, with $A=D$. \revision{(b) Equivalent periodic lensguide of the resonator shown in (a). $A_1B_1C_1D_1$ is the one-way resonator matrix from the left to the right plane end mirrors. The round-trip resonator matrix $ABCD$ is obtained as the ordered product of the two one-way matrices $A_1B_1C_1D_1$ and $D_1B_1C_1A_1$ \cite{r28}.} (c) Optical resonator used to simulate non-Hermitian anharmonic motion of a Gaussian beam. The resonator is stable for $L<f$.}
%\end{figure}

%\acknowledgments

\end{document}